\documentstyle[12pt,epsf]{article}

\newcommand\myappendixsection{\setcounter{equation}{0} \section}
\def\lrang#1{\left\langle#1\right\rangle}
\def\beql#1{\begin{equation}\label{#1}}
\def\eeq{\end{equation}}
\def\beq{\begin{equation}}
\def\beeq{\begin{eqnarray}} \def\eeeq{\end{eqnarray}}
\newcommand\mysection{\setcounter{equation}{0}\section}
\renewcommand{\theequation}{\thesection.\arabic{equation}}
\newcounter{hran} \renewcommand{\thehran}{\thesection.\arabic{hran}}

\def\bmini{\setcounter{hran}{\value{equation}}
  \refstepcounter{hran}\setcounter{equation}{0}
  \renewcommand{\theequation}{\thehran\alph{equation}}\begin{eqnarray}}

\def\bminiG#1{\setcounter{hran}{\value{equation}}
\refstepcounter{hran}\setcounter{equation}{-1}
\renewcommand{\theequation}{\thehran\alph{equation}}
\refstepcounter{equation}\label{#1}\begin{eqnarray}}

\def\emini{\end{eqnarray}\relax\setcounter{equation}{\value{hran}}\renewcommand{\theequation}{\thesection.\arabic{equation}}}

\def\fm{\,{\rm fm}}
\def\eV{{\rm e\kern-0.12em V}}            \def\MeV{{\rm M}\eV}
 \def\GeV{{\rm G}\eV} 
\def\lsim{\raise0.3ex\hbox{$<$\kern-0.75em\raise-1.1ex\hbox{$\sim$}}}
\def\gsim{\raise0.3ex\hbox{$>$\kern-0.75em\raise-1.1ex\hbox{$\sim$}}}
 \def\cite#1{[\ref{#1}]}
 \def\citd#1#2{[\ref{#1},\ref{#2}]}
 \def\citt#1#2#3{[\ref{#1},\ref{#2},\ref{#3}]}

\def\np#1#2#3{{\em Nucl.~Phys.}~\underline{B#1} (19#3) #2}
\def\pl#1#2#3{{\em Phys.~Lett.}~\underline{#1B} (19#3) #2}
\def\pr#1#2#3{{\em Phys.~Rev.}~\underline{#1} (19#3) #2}

\def\prl#1#2#3{{\em Phys.Rev.Lett.}~\underline{#1} (19#3) #2}
\def\sjnp#1#2#3{{\em Sov.J.Nucl.Phys.}~\underline{#1} (19#3) #2}

\begin{document}
\begin{flushright}CU-TP-760 \\
   BI-TP 96/26 \\
   LPTHE-Orsay 96/61
\end{flushright}
\vskip 50pt
\begin{center}
\begin{title}
\title{\Large \bf Radiative energy loss and $p_\perp$-broadening
of high energy partons in nuclei}
\end{title}

\vskip 24pt
\begin{author}
\author{R. Baier$^1$, Yu. L. Dokshitzer$^2$, A.H.
Mueller$^3$, S. Peign\'e$^4$, and D. Schiff$^4$}
\end{author}
\end{center}
\vskip 50pt

\begin{center}
$^1$Fakult\"at f\"ur Physik, Universit\"at Bielefeld, D-33501
Bielefeld, Germany\\

$^2$
{INFN, Sezione di Milano, Italy$^\dagger$}\\

$^3$Physics Department,$^{*}$ Columbia University, New York, N.Y.
10027, USA\\

$^4$LPTHE$^{\ddagger}$, Universit\'e
Paris-Sud, B\^atiment 211, F-91405 Orsay, France 
\end{center}

\vfill
\begin{abstract}
The medium-induced $p_\perp$-broadening and induced gluon radiation spectrum of
a high energy quark or gluon traversing a large nucleus is studied.  Multiple
scattering of the high energy parton in the nucleus is treated in the Glauber
approximation.  
We show that $-{dE/ dz}$, the radiative energy loss of the parton per unit length,
grows as $L$, the length of the nuclear matter, as does 
the characteristic transverse momentum squared of the parton $p_{\perp W}^2$.  
We find $-{dE/ dz} = {\textstyle 1\over 8} {\alpha_sN_c}  
p_{\perp W}^2$
holds independent of the details of the parton-nucleon scatterings so long 
as  $L$  is large. 
Numerical estimates suggest that $p_\perp$-broadening and energy loss may be
significantly enhanced in hot matter as compared to cold matter, thus making
the study of such quantities a possible signal for quark-gluon plasma formation.

\end{abstract}

\vfill

\noindent$^*$Supported in part by the U.S. Department of Energy under Grant
DE-FG02-94ER-40819\\
\noindent$\dagger$ Permanent address: Petersburg Nuclear Physics Inst.,
Gatchina 188350, St. Petersburg, Russia\\
\noindent$\ddagger$Laboratoire associ\'e au Centre de la Recherche Scientifique
- URA D0063
 \newpage

\mysection{\bf Introduction}

In a recent paper \cite{BDMPSII}, 
hereinafter referred to as BDMPS II or simply as II, we
investigated the problem of induced gluon radiation and radiative energy loss
of a high energy quark or gluon traversing hot QCD matter of finite volume. 
This induced radiation may be an important tool for studying the physics of
relativistic heavy ion collisions.  In particular, contrasting hot and cold
matter results may provide a signal for quark-gluon plasma formation.  The
induced gluon radiation may be measurable by studying the medium dependence of
the characteristics of high-$p_\perp$ jet and hadron production.

The present paper is devoted to the study of radiative energy loss and jet
$p_\perp$-broadening in nuclear matter.   We shall show that these two
phenomena are related in a model-independent way.  Scatterings encountered by a
fast parton in a nuclear medium may be described within a diffusion picture. 
The jet $p_\perp$-broadening is determined by a classical diffusion equation. 
Energy loss is described in a picture where the fast parton propagation is
treated classically whereas the induced gluon radiation is a quantum
mechanical phenomenon.  Interference effects play a crucial role in the
production of gluons which determine the induced radiation spectrum.

The Gyulassy-Wang model \cite{GW} used in II to describe multiple scatterings in hot
matter is here replaced by the Glauber model \cite{Glauber} 
of multiple scattering in which
parton-nucleus scattering is given in terms of independent scatterings off the
nucleons making up the nucleus.

The result which we find for radiative energy loss is similar to that for hot
matter, namely that the total energy loss is proportional to the square of the
length of the traversed nuclear matter, $L$. The coefficient of $L^2$ may be
estimated from other, nonperturbative but measurable, quantities such as the
small-$x$ gluon distribution of the nucleon or, as discussed by Luo, Qiu and
Sterman in \cite{LQS}, 
the transverse momentum which a jet, or dijet, receives as it passes
through a nucleus.  In this latter case, we find an interesting general
relation between the radiative energy loss per length and 
the $p_\perp$-broadening (defined as the characteristic width $p_{\perp W}^2$
of the transverse momentum distribution) of a high energy parton passing
through a nucleus,
$-{dE/ dz} = \frac18 {\alpha_sN_c} p_{\perp W}^2$.  
Numerical estimates suggest possibly large differences for $p_\perp$-broadening 
and energy loss for partons in hot matter as compared to nuclear matter.  

The outline of the paper is as follows. In section 2 the basic equations are
established for jet $p_\perp$-broadening and for the induced gluon radiation
spectrum.  These equations are solved in section 3 to logarithmic accuracy. 
Parameters appearing in the solutions are related to (other) phenomenological
quantities, such as the gluon distribution of the nucleon, in section 4.  Nu\-me\-ri\-cal
estimates are given in section 5.  In Appendix A, our basic equation for jet
broadening is derived in detail.  In Appendix B we relate the elementary parton
nucleon scatterings, as they are used for $p_\perp$-broadening and  radiative
energy loss, to the gluon distribution of the nucleon.

\mysection{The equations}

In this section we give the basic equations for jet transverse momentum
broadening and induced gluon radiation for a high energy jet traversing cold
nuclear matter.  The equation for jet broadening is straightforward to write
down and represents diffusion, in transverse momentum, due to independent
multiple scatterings of the jet as it encounters the various nucleons of the
nucleus.  The equation for induced energy loss has exactly the same form as
found in II.  In each case the fundamental assumption is that a parton is multiply
scattered off uncorrelated scattering centres in the material.  That is, we
suppose that the Glauber picture of multiple scattering holds for a high
energy parton passing through a nucleus.  At extremely high energies the Glauber
approximation breaks down due to coherent effects in inelastic particle
production  \cite{Gribov}.  
We limit the energies which we consider in order to avoid this
complication.  It is an important goal for the future to better understand how
severe these limitations are in practice.

\subsection{The  basic  elements  of  the  equations}

We formulate the multiple scattering in close analogy with the procedure 
used in \cite{BDPS} within the Gyulassy-Wang model.  
Thus, we first need to find the cold matter quantity
corresponding to the single scattering probability distribution.


In jet $p_\perp$-broadening the high energy parton receives a large 
transverse momentum due to the many scatterings in passing through a large nucleus.  
This means that in the individual parton-nucleon collisions all gluon couplings 
to the high energy parton occur with a small impact parameter separation between 
the parton in the amplitude and in the complex conjugate amplitude.
This, perhaps surprising, result can be seen by referring below to (\ref{B.huit}) .
If one inserts a complete set of scattering states between the two $F^a_{+i}$'s
in the right-hand side of that equation, the cross section is given explicitly in terms
of a production amplitude times a complex conjugate amplitude.
The amplitude $F^a_{+i}$ is evaluated at $\vec{y}=0$ while in the complex conjugated
amplitude $\vec{y}\,^2=B^2/\mu^2 $ with  $\mu$ a scale to be defined shortly.
With such small impact parameter separations, $B^2\propto L^{-1}$, 
the one-gluon approximation becomes accurate.  
(The argument of the parton-gluon coupling is just the impact parameter separation.)  

In general, we allow breakup of the struck nucleon in the elementary
parton-nucleon collision.  
The basic high energy quark-nucleon scattering amplitude is shown in Fig.1 and the 
corresponding cross section is denoted by ${d\sigma/ d^2\vec q}$. 

\begin{figure}[h]
\centering
\leavevmode
\epsfxsize 6cm
\epsffile{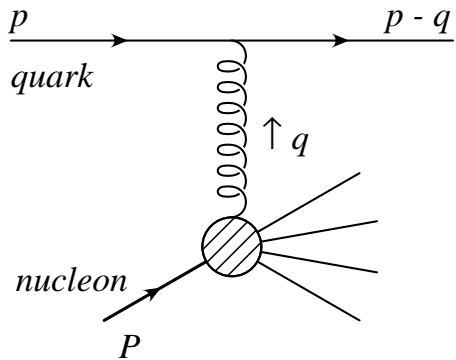}
\caption{\it{Basic quark-nucleon scattering amplitude.}}
\label{fig1}
\end{figure}

\noindent A crucial assumption of the Glauber model we are using to describe multiple 
scattering 
is the fact that successive scatterings are independent.  
That is, the collision process at a given centre does not depend on the collisions 
which occur before or after.  
If the particles produced in the elementary collision  in Fig.1 have a formation time 
smaller than the average time between two scatterings  (the mean free path $\lambda$) 
then the independent scattering picture is clear.  
If that formation time is longer one expects corrections to independent 
scattering \cite{Gribov}.   
In the following sections and Appendix B  we shall relate an elementary scattering to the
gluon distribution  of the nucleon.  

An estimate of the size of the collective effects which undermine the Glauber 
model is the strength of shadowing corrections for the gluon distribution of a nucleus. 
Phenomenologically, these corrections are always less than a factor of  2 
even for quite small values of $x$ \citd{Frankfurt}{Eskola}.  
Given this uncertainty, we expect the independent scattering picture to be 
a reasonable framework for studying jet $p_\perp$-broadening and the induced 
gluon spectrum for high energy partons. 

Because ${d\sigma/ d^2\vec q}$ obeys diffractive kinematics, the
four-momentum transfer squared, $q^2$, is equal to ${\vec q\ ^2}$ the
two-dimensional momentum transfer squared.
It is 
then 
convenient to define
\beql{2.1}
\sigma = \int {d\sigma\over d^2\vec q}\  d^2\vec  q
\eeq
as the total elementary cross section.  The reader may worry
about the meaning of $\sigma$ since the integration over small transverse
momenta is not under perturbative control.  We may imagine that the integral on
the right-hand side of (\ref{2.1}) has an infrared cutoff.  The parton-nucleon
cross section cannot be reliably calculated in perturbative QCD.  
However, we shall show that the cutoff dependence disappears 
(in the logarithmic approximation)
when one calculates quantities insensitive to small momentum transfer scattering
such as jet $p_\perp$-broadening and induced gluon radiation.
This cutoff  independence is due to a cancellation of 
long-distance effects between the real inelastic production shown in Fig.1 and
forward elastic scattering, a virtual contribution.  
This cancellation is required by the probability conserving formalism which we use 
to describe multiple scattering.
Any divergence in real production must be compensated by a corresponding virtual
contribution (the no-scattering term on the right-hand side of (\ref{2.7})).
This virtual contribution
in our formalism shows up in the (cutoff dependent) mean free path for the quark
given by
\beql{2.2}
\lambda = [\rho\sigma]^{-1} \>,
\eeq
with $\rho$ the nuclear matter density.  

In order to deal with dimensionless variables, and functions, it is convenient
to introduce a (somewhat arbitrary) scale $\mu^2$, with $\mu$ representing a
$typical$  momentum transfer to the quark in a quark-nucleon collision.  
Then $\vec Q = \vec q/\mu$ and $\vec U = \vec p/\mu$ are the dimensionless
transverse momentum transfer and transverse momentum, respectively.  
The probability distribution for the scattered quark is given by
\beql{2.3}
V(Q^2) = {1\over \pi\sigma}\ \ {d\sigma\over dQ^2} \>,
\eeq
where we note that $\int d^2\vec Q V(Q^2) = 1$.  
The ``potential'' $V$,  defined by
(\ref{2.3}), will play exactly the same role in cold matter calculations as did a
similar quantity, also called  $V$,  for hot matter \cite{BDPS}.

As was done earlier for hot matter it is convenient to define the multiple
scattering of a high energy parton in terms of the probability distribution for
a single scattering, $V(Q^2)$, and the probability distribution for distances
$\Delta$ between successive scatterings \citd{GW}{BDPS},
\beql{2.4}
P(\Delta) d\Delta = e^{-\Delta/\lambda} \frac{d\Delta}{\lambda} \> .
\eeq

\subsection{Equation for parton 
$p_\perp$-broadening}

In calculating the transverse momentum given to a high energy parton by multiple
scattering with the nucleons of a nucleus it is not necessary, in the leading
approximation, to include the transverse momentum given to the parton by induced
gluon emission since the emission is formally of higher order in ${\alpha}_{s}$.   
Thus the transverse momentum given to the jet in passing through a nucleus
comes directly from the scattering of the projectile parton
off the nucleons in the nucleus.

Suppose a high-energy parton is produced in the nucleus in a hard collision at $z=0$ with 
an initial transverse momentum distribution $f_0(U^2)$ with
\beql{2.5}
\int d^2\vec U f_0(U^2) = 1\>.
\eeq
(If the parton initially has no transverse momentum, then $f_0(U^2) =
{1\over \pi} \delta(U^2)$.)  In terms of the scaled variable
\beql{2.6}
t = \frac{z}{\lambda_R} \>, \quad \lambda_R=\lambda{C_F\over C_R}
\eeq
for a parton of colour representation  $R$,  the transverse momentum
distribution at $z=t \lambda_R$ is
\beql{2.7}
f(U^2,t) = f_0(U^2) e^{-t} + \int_0^t dt^\prime e^{-(t-t^\prime)}\int d^2\vec
Q f\left((\vec U-\vec Q)^2, t^\prime\right)\,V(Q^2) \>.
\eeq
The distribution $f(U^2,t)$ satisfies 
the normalization condition 
(\ref{2.5}) for all $t$.  
We derive (\ref{2.7}) using a multiple scattering formalism in Appendix A.  
The first term on the right-hand side of (\ref{2.7}) corresponds to no interaction 
with the nuclear medium where the $e^{-t}$ represents the probability that 
no interactions have occurred in a distance $z=\lambda_Rt$.  
The second term on the right-hand side of (\ref{2.7}) corresponds to a final scattering 
at $t^\prime$ before the parton reaches  $t$  with transverse momentum $\vec U$.  
One can recast (\ref{2.7}) as a differential equation by taking a ``time'' 
derivative on both sides of that equation.  
One finds
\beql{2.8}{\partial f(U^2,t)\over \partial t} = - f(U^2,t) + \int d^2 \vec Q f((\vec
U-\vec Q)^2,t) V(Q^2) \>.
\eeq
A different, but equivalent, form of (\ref{2.8}) for  $f$,  which makes its
interpretation as a kinetic master equation more transparent, is
\beql{2.9}
{\partial f(U^2,t)\over \partial t} = 
- \int f(U^2,t)          V((\vec U-\vec U^\prime)^2) d^2\vec U^\prime 
+ \int f(U^{\prime 2},t) V((\vec U^\prime-\vec U)^2) d^2\vec U^\prime \>.
\eeq
When partons pass through the distance   $dt$  the first term on the
right-hand side (loss term) accounts for partons which are scattered {\em out}\/ of
the direction $\vec U$, $\vec U \to \vec U^\prime$, weighted by the (normalized)
scattering  cross section $V$.  The second term (gain term) counts those partons
which are scattered $into$ the direction $\vec U$ from all other directions
$\vec U^\prime, \vec U^\prime \to \vec U$.  
One can diagonalize (\ref{2.8}) by defining
\bminiG{2.10}
\label{2.10a}
\tilde f(B^2,t) = \int d^2\vec U e^{-i\vec B\cdot \vec U} f(U^2,t) \>,
\eeeq
and
\beeq
\label{2.10b}
\tilde V(B^2) = \int d^2\vec Q  e^{-i\vec B\cdot \vec Q}V(Q^2)\>,
\emini 
in which case
\beql{2.11}
{\partial \tilde f(B^2,t)\over \partial t} = -{1\over 4}
B^2\tilde v(B^2)\tilde f(B^2,t)\>,
\eeq
where
\beql{2.12}
\tilde v(B^2) = {4\over B^2}(1-\tilde V(B^2))\>.
\eeq
Note that $\tilde V(0)=1$ because of our
normalization of $V(Q^2)$ as a probability distribution.

Instead of searching for an exact solution one often approximates (\ref{2.8}),
in case of soft ``potentials'', 
by expanding $f((\vec U-\vec Q)^2,t)$ inside the integral through second order in
$\vec Q$.  This leads to the diffusion equation \cite{Migdal}
\beql{2.13}
{\partial f(U^2,t)\over \partial t} = {1\over 4}\, {\hat q}_{_U} 
\bigtriangledown_U^2 f(U^2,t)\>,
\eeq
where the ``transport coefficient''  $\hat q_{_U}$ is defined by
\beql{2.14}
\hat q_{_U} = \int d^2\vec Q\, Q^2V(Q^2) \>.
\eeq
The transport coefficient with respect to the longitudinal
variable  $z$ (see (\ref{2.6})) and the transverse momentum $\vec q$ is then given by

\beql{2.15}
\hat q = {\mu^2\over \lambda_R} \hat  q_{_U}\>.
\eeq
In cases where the integral in (\ref{2.14}) exists one finds that
\beql{2.16}
\hat q_{_U} = \tilde v(0)\>.
\eeq
For example, if  $V$  were gaussian, $V(Q^2) = {1\over \pi} e^{-Q^2}$,
(\ref{2.7}) and (\ref{2.14}) would be completely equivalent to standard 
treatments of parton transverse momentum broadening in nuclei \cite{BBL} 
with $\mu^2$ 
the average transverse momentum in an elementary collision.

However, in QCD (\ref{2.14}) diverges logarithmically since $V(Q^2) \propto 1/Q^4$
at large $Q^2$. 
Correspondingly, $\tilde v(B^2)$ has no finite limit for $B^2 \to 0$. 
Nevertheless, $\tilde v(B^2)$ may be related to the {\em first  moment}\/
of $V(Q^2)$ in a logarithmic approximation.  
To see this we use (\ref{2.10b}) and (\ref{2.12}) to write
\beql{2.17}
\tilde v(B^2) = {4\over B^2} \int d^2\vec Q \left(1-e^{-i\vec B\cdot \vec Q}\right)
V(Q^2)\>.
\eeq
Doing the angular integral 
gives
\beql{2.18}
\tilde v(B^2) = {4\pi\over B^2} \int_0^\infty dQ^2(1-J_0(BQ))\, V(Q^2) \>.
\eeq
This integral is convergent in the ultraviolet.  For $BQ \ll 1$, 
we have
$1\!-\!J_0(BQ) \approx {\textstyle 1\over 4} B^2Q^2$  so that there is a logarithmic 
region of integration in (\ref{2.18}) extending up to $Q^2 \approx 1/B^2$.  
Thus, in the logarithmic approximation, for small $B^2$ 
\beql{2.19}
\tilde v(B^2) \approx \pi \int_0^{1/B^2} dQ^2 V(Q^2) Q^2 \>.
\eeq


\subsection{Equation for the induced energy spectrum}

Because of the close correspondence between the multiple scattering in hot
matter, as discussed in II, and the present circumstance we need not rederive
the equation for the induced radiation spectrum.  We may simply take over
equations (4.16), (4.24) and (5.1) of that paper which read
\beql{2.20}
{\partial\over \partial \tau} \vec f(\vec U,\tau) = - (1-i\tilde \kappa U^2) \vec
f(\vec U,\tau) + \int d^2 \vec{Q} \vec f(\vec U-\vec Q, \tau) V(Q^2)\>, 
\eeq
\beql{2.21}
{\omega dI\over d\omega dz} = {3\alpha_sC_R\over 2\pi^2\lambda_g}\> 2  {\rm Re}
\, \left\{ 
i \int_0^{\tau_0} d\tau\left(1-\frac{\tau}{\tau_0}\right) \int {d^2\vec B\over
2\pi}\ {1-\tilde V(B^2)\over B^2} \vec B\cdot 
\widetilde{\vec{f_{\ }}}(\vec B,\tau)\bigg\vert_{\tilde \kappa}^{\tilde \kappa =0} 
\>\right\},
\eeq
and
\beql{2.22}
i{\partial\over \partial\tau} \widetilde{\vec{f_{\ }}} (\vec B,\tau)
=\left[\,\tilde \kappa\nabla_B^2-{i\over 4}B^2\tilde v(B^2)\,\right]
\widetilde{\vec{f_{\ }}}(\vec B,\tau)\,;
\quad 
\widetilde{\vec{f_{\ }}}(\vec B,0) = - {i\pi\over 2}\vec B\ \tilde v(B^2) \>,
\eeq
respectively.  
In (\ref{2.20})--(\ref{2.22})
\beql{2.23}
\tau = z/\tilde \lambda, \quad  \tau_0=L/\tilde \lambda \>; \qquad 
\tilde \kappa = {2C_F\over N_c} \kappa\>,\quad  
\tilde \lambda = {2C_F\over N_c} \lambda = 2\lambda_g \>.
\eeq
Equations (\ref{2.21}) and (\ref{2.22}) give the induced gluon radiative energy
spectrum for high energy partons having colour representation  $R$.  
The importance of the parameter $\kappa = {\lambda \mu^2/ 2\omega}$ 
has been discussed in~\cite{BDPS}.

\mysection{Energy loss and jet broadening}

In this section, we give the solutions to (\ref{2.20}) -- (\ref{2.22}).  
We then compare the resulting formulas for ${dE/ dz}$ and $p_{\perp W}^2$ 
and note a simple relationship between these two quantities.

\subsection{Jet  transverse  momentum  broadening}

It is straightforward to solve (\ref{2.11}) as
\beql{3.1}
\tilde f(B^2,t) = \tilde f(B^2,0) \exp\left\{-{1\over 4} B^2 \tilde v(B^2) t\right\}\>.
\eeq
In this paper we only consider media whose length  $L$  obeys
$L/\lambda \gg 1$.  Thus we need only consider  $t$  large in which case
only small values of $B^2$ give a result which is not exponentially small.  
If $\tilde v(0)$ were finite we could choose $\mu^2$ to be 
the mean momentum transfer squared per collision
so that $\tilde v(0)\equiv 1$.  
In this case the transverse momentum distribution, $f(U^2,t)$, would be a gaussian
in $U$ with {\em mean transverse momentum squared}\/ $\lrang{p_{\perp}^2}=\mu^2 
t$.
However, we expect a logarithmic dependence of $\tilde v(B^2)$ for small $B^2$ 
because of the approximate scale invariance of QCD.  
We suppose that $B^2{\partial\over \partial B^2} \ln \tilde v(B^2) \ll 1$ 
for small $B^2$ in which case it is clear from (\ref{3.1}) that the values 
of $B^2$ for which $\tilde f(B^2,t)$ is not small are
\beql{3.2}
B^2 \leq {4\over t\tilde v(1/t)} \>.
\eeq
Using $\tilde f(0,0) = 1$, coming from the normalization of $f(U^2,t)$ as
a probability distribution in transverse momenta, we get
\beql{3.3}
\tilde f(B^2,t) \approx \exp\left\{-{1\over 4} B^2t\tilde v({1/ t})\right\}\>.
\eeq
Transforming back to $\vec U$-space gives
\beql{3.4}
f^{(G)}(U^2,t) \approx {1\over \pi\,t \tilde v(1/t)} 
\exp\left\{-{U^2\over  t\,\tilde v({1/ t})} \right\}\>.
\eeq
In neglecting the $B^2=0$ singularity of $\tilde{v}$ in (\ref{3.4})
we have lost the large-$U^2$ tail of $f(U^2,t)$ which is
$$
    f(U^2,t) \>\simeq\> V(U^2)\,t\>\sim\> \frac{t}{U^4}\>.
$$
The gaussian approximation (\ref{3.4})
is valid up to $U^2$-values where the exponential is as small as $1/t\ll1$.
This means that to evaluate the {\em characteristic width}\/ of the distribution one can
use (\ref{3.4}).
Defining $p_{\perp W}^2$ as the value at which the distribution falls to $1/e$ of its
peak value, we have 
\beql{3.5}
p_{\perp W}^2 = \mu^2 \int d^2\vec U \,\vec U^2 f^{(G)}(U^2, L/\lambda_R)\>,
\eeq
which results in
\beql{3.6}
p_{\perp W}^2 = {\mu^2\over \lambda_R} \tilde v(\lambda_R/L)\, L\>.
\eeq
The linear growth of typical transverse momenta squared with $L$ is expected and has 
been
used for some time to discuss  $p_\perp$-broadening of high energy
partons in nuclei \citt{BBL}{Alde}{LQS}.  
  
The individual parts on the right-hand side of (\ref{3.6}) which multiply  $L$ 
are not uniquely defined.  
Nevertheless, we note that the quantity ${\mu^2\over \lambda_R} \tilde v(\lambda_R/L)$ 
is independent of the scale $\mu^2$ and of any infrared cutoff we may have used 
to define $\sigma.$  
Indeed $\tilde v$ scales as $1/\mu^2$ so that $\mu^2 \tilde v$ is independent of $\mu^2$.  
More importantly, the factors of $\sigma$ cancel in $\tilde v/\lambda$, and the dangerous 
small-$\vec Q$ region of integration is suppressed for small $\vec B$ in taking the
difference indicated in (\ref{2.17}) leaving the cutoff only as a scale for
logarithmic dependence.  
Indeed, from (\ref{2.19}) 
\beql{3.7}
{\tilde v(B^2)\over \lambda} \approx \rho \int^{1/B^2} dQ^2
{d\sigma\over dQ^2}Q^2 \>,
\eeq
which explicitly shows the suppression of long-distance contributions.

It is worth mentioning that the basic evolution equations (\ref{2.11}) and (\ref{2.22}),
and thus their solutions, are also cutoff independent for
large  $z$ if we substitute $t=z/\lambda_R$ everywhere in the equations and
view $\tilde f$  as a function of the physical variable  $z$ rather than $t$.  
We prefer to continue working with the variable $t$, however, as the
correspondence with the procedure used for hot matter is more transparent in
this variable.

\subsection{Energy  loss}

Since (\ref{2.21}) and (\ref{2.22}) are equations identical to those encountered in II we
may immediately write down the solution to (\ref{2.22}) as
\beql{3.8}
\widetilde{\vec{f_{\ }}}(\vec B,\tau)\  = {2\pi i\vec B\over B^2}\ {\partial\over \partial
\tau} \exp\left\{-{i\over 2} m\  \omega_0B^2 \tan  \omega_0\tau \right\} \>,
\eeq
and we may write (\ref{2.21}) as
\beql{3.9}
{\omega dI\over d\omega dz} = {6\alpha_sC_R\over \pi L} \ln\left|{\sin
\omega_0\tau_0\over \omega_0\tau_0}\right| \>,
\eeq
where
\beql{3.10}
m = - {1\over 2\tilde \kappa}\>,\quad  \omega_0={\sqrt{i\tilde \kappa \tilde v}}\>.
\eeq
The argument of $\tilde v$ is $B^2$ in (\ref{3.8}) and
$\sqrt{\tilde \kappa/\ln(1/\tilde \kappa)}$ in (\ref{3.9}).  
Exactly as in II one may give an analytic expression for ${dE/ dz}$, 
the energy loss per unit length, as
\beql{3.11}
-{dE\over dz} = {\alpha_sC_R\over 8}\ {\mu^2\over \lambda_g}\
\tilde v(\tilde \lambda/L)\ L = {\alpha_sN_c\over 8}\ {\mu^2\over \lambda_R}\
\tilde v(\tilde \lambda/L) L \>,
\eeq
and the total energy loss  as
\beql{3.12}
-\Delta E = {\alpha_sC_R\over 8}\ {\mu^2\over \lambda_g} \tilde v(\tilde \lambda/L)\
L^2 \>.
\eeq
In the above, $\alpha_s=\alpha_s(k^2)$ is evaluated at a large momentum scale 
$k^2$ proportional to $L$.  Indeed, as shown in II, the effective value of the impact
parameter $B^2$ is of order $\tilde \lambda/L.$

\subsection{Relationship  between  energy  loss  and  $p_\perp$-broadening}

Comparing (\ref{3.6}) and (\ref{3.11}) we note that
\beql{3.13}
-{dE\over dz} = {\alpha_sN_c\over 8}\ p_{\perp W}^2
\eeq
relating the energy loss per unit length with the typical transverse
momentum squared that a jet receives in passing through a length $L$ of
nuclear matter.  
It is interesting that the
relationship between these two quantities is completely independent of the
dynamics of the multiple scattering, at least in the framework of our approach
to these problems.  
Thus (\ref{3.13}) holds equally well in finite length hot matter
as well as cold matter.  
Perhaps even more surprising the coefficient relating
$p_{\perp W}^2$ to ${dE/ dz}$ is independent of the nature of the high
energy parton passing through the hot or cold matter.  
Equation (\ref{3.13}) makes more precise the bound 
$-{dE/ dz} \leq {1\over 2} p_{\perp W}^2$ 
suggested sometime ago by Brodsky and Hoyer \cite{BH}
on the basis of the uncertainty relation.

\mysection{Determining ${\mu^2\over \lambda} \tilde v$, estimates}

In this section, we shall discuss how ${\mu^2\over \lambda}\tilde v$ can be
determined phenomenologically and we shall give a crude estimate of this
quantity in terms of the nucleon gluon distribution. 

\subsection{Determining\  ${\mu^2\over \lambda} \tilde v$  phenomenologically}

The broadening of the transverse momentum spectrum of high energy partons in
nuclei has been widely discussed in the literature.  
In $\mu$-pair and $J/\psi$-production in high energy proton-nucleus collisions 
it is the incoming high energy quark (or gluon) which multiply scatters in the nucleus
and in so doing broadens the $\mu$-pair and $J/\psi$ transverse momentum
spectra as compared to the corresponding production in nucleon-nucleon
collisions.  
Transverse momentum broadening has also been studied for outgoing
jets produced in high energy proton-nucleus and photon-nucleus collisions. 

The phenomenological situation is somewhat confused at the moment.  
It would appear that {\em outgoing}\/ partons receive much more transverse momentum 
due
to multiple scattering in a nucleus \cite{FC} than do {\em incoming}\/ 
partons \cite{Alde}.  
Theoretically, this difference is not understood \citd{LQS}{LR}.  
It is not the purpose of this section to give a comprehensive study of 
the phenomenology of multiple scattering of high energy partons in nuclei 
but rather to show how ${\mu^2\over \lambda}\tilde v$
can be determined, so we limit ourselves here to relating our approach to that
of Luo, Qiu and Sterman \cite{LQS} who studied the $p_\perp$-broadening of dijets
produced in photon-nucleus collisions at Fermilab.

Luo, Qiu and Sterman consider a single rescattering in the nucleus of dijets
produced in high energy photon-nucleus collisions.  
They find that the transverse momentum squared given to the dijet 
by summing over all single rescatterings in the nucleus is (see equation (21) in 
\cite{LQS}):
\beql{4.1}
\lrang{k_T^2(R)} = {4\over 3} \pi^2\alpha_s(Q^2)\, A^{1/3}\, \lambda_{LQS}^2 \>.
\eeq
Here  $Q$  is the relative transverse momentum of the two jets making
up the dijet,   $A$ is the atomic number of the nucleus and
$\lambda_{LQS}^2$ is a dimensional parameter which characterizes the momentum 
transfer
squared in a single collision.  
An expression for $\lambda_{LQS}^2$ is given in \cite{LQS} in terms of 
a new 
QCD nuclear matrix element, but the actual determination
of $\lambda_{LQS}^2$ is made by comparing (\ref{4.1}) to experiment.  
A value $\lambda_{LQS}^2 \approx 0.05-0.1 {\GeV}^2$ is found.

We may compare (\ref{3.6}) to (\ref{4.1}) by noting the following: 

(i)  In the Luo, Qiu and Sterman calculation, 
the dijet consists of a high transverse momentum quark
and a gluon with the quark and gluon transverse momenta nearly balancing. 
It is the nuclear contribution to the imbalance, namely, the transverse
momentum of the whole dijet, which is expressed in (\ref{4.1}).  
Since the quark and gluon have very high transverse momentum they form 
a (colour triplet) system which is compact in transverse coordinate space.  
Thus, in a relatively soft rescattering with a nucleon in the nucleus this system should 
act like a single quark and this indeed is the case in the calculation done in \cite{LQS}.  
We may then expect that our calculation of jet broadening of a single parton
should apply to the dijet system.  

(ii)  Luo, Qiu and Sterman only consider a
single rescattering while we have summed arbitrary numbers of rescatterings in
arriving at (\ref{3.6}) from (\ref{2.7}).  
However, in a multiple scattering, where the
scatterers are uncorrelated, the resulting transverse momentum squared is just
the sum of the transverse momenta squared of the individual scatterings.  
In \cite{LQS} only a single scattering is explicitly done but (\ref{4.1}) results from
summing over all possible single scatterings.  
Thus (\ref{3.6}) and (\ref{4.1}) should be directly comparable.
However, here we do sidestep the difficult issue as to exactly which 
$\lrang{k_T^2}$-average has been done experimentally in evaluating 
$\lambda^2_{LQS}$
and how close that average is to our $p_{\perp W}^2$.

The last piece of information we need 
is to relate $L$  to the nuclear radius $R$. 
Since the dijet is produced in a hard collision it is produced uniformly 
in the volume of the nucleus. 
After being produced the average length of material that the dijet passes through is 
$L = {3\over 4} R$.  
With this identification of $L$ 
we may equate (\ref{3.6}) to (\ref{4.1}) and obtain
\beql{4.2}
{\mu^2\over \lambda} \tilde v(\lambda/L) = {16\over 9} \pi^2\alpha_s(Q^2)\,
{A^{1/3}\over R}\, \lambda_{LQS}^2 \>.
\eeq
We shall give a numerical estimate of ${\mu^2\over \lambda} \tilde v$ from
(\ref{4.2}) a little later on.

\subsection{Relating ${\mu^2\over \lambda} \tilde v$ to the
gluon distribution}


One can get a crude relation between ${\mu^2\over \lambda}\tilde v$  and the
nucleon gluon distribution \cite{Levin} and that is our purpose in this section.  
Indeed, Luo, Qiu and Sterman relate their parameter $\lambda_{LQS}^2$  to the average
colour field strength squared that a parton {\em sees}\/ as it passes through a
nucleon in the nucleus.  
In order to relate ${\mu^2\over \lambda}\tilde v$ to the
gluon distribution we use (\ref{2.19}) which gives
\beql{4.3}
\tilde v(B^2) = {1\over \sigma_R} \int_0^{1/B^2} dQ^2\, Q^2 {d\sigma_R\over dQ^2}
= {1\over \mu^2\sigma_R} \int_0^{\mu^2/B^2} d q^2\, q^2 {d\sigma_R\over d
q^2}\>,
\eeq
where now ${d\sigma_R/ d Q^2}$ refers to the scattering of a
parton of colour representation $R$ with a nucleon.  
But in the one-gluon exchange approximation
\beql{4.4}
\int_0^{\mu^2/B^2} d q^2 q^2 {d\sigma_R\over d q^2} = \frac{4\pi^2\alpha_s 
C_R}{N_c^2-1}
\, x G(x, \mu^2/B^2) \>.
\eeq
(We postpone until later on the prickly question of what value of $x$
we intend to use in (\ref{4.4}).  
We do note however that in the region that (\ref{4.4})
will be used, a region of small but not too small  $x$,  the quantity 
$x G(x,\mu^2/B^2)$ should have little dependence on $x$.)  
The simplest way to see that (\ref{4.4}) should be true is to consider 
the case where the  target and jet are both quarks.  
In this case
\beql{4.5}
{d\sigma\over d q^2}\ =\ {2\pi C_F\alpha_s^2\over N_c\,(q^2)^2}
\eeq
to lowest order in $\alpha_s$ while
\beql{4.6}
x G_q(x, \mu^2/B^2) = {\alpha_sC_F\over \pi}\ \ln \frac{\mu^2}{B^2} \>.
\eeq
Using (\ref{4.5}) in the left-hand side of (\ref{4.4}) gives the right-hand side
of that equation after using (\ref{4.6}), thus fixing the 
${2\pi^2\alpha_s/N_c}$ factor on the right-hand side of (\ref{4.4}), for a quark jet, 
and a factor ${4\pi^2\alpha_sC_R/(N_c^2-1)}$ for a jet of colour representation $R$.  
(A more complete, and perhaps more satisfactory, derivation of (\ref{4.4}) is given in
Appendix~B.)  
The $\alpha_s$ in (\ref{4.4}) should be evaluated at the scale
$\mu^2/B^2$ as $B/\mu$ is the short distance over which the scattering
occurs.  
Using (\ref{4.4}) in (\ref{4.3}) along with $\lambda_R = [\rho \sigma_R]^{-1}$
gives
\beql{4.7}
{\mu^2\over \lambda_R} \tilde v(B^2) = {4\pi^2\alpha_sC_R\over N_c^2-1}\, \rho\,  
xG(x,\mu^2/B^2) \>.
\eeq

\mysection{Some numerical estimates}

The basic results for $p_\perp-$ broadening and energy loss are given in (\ref{3.6})
and in (\ref{3.11}), (\ref{3.12}) respectively.  The parameter controlling these
quantities is ${\mu^2\over \lambda}\ \tilde v$, for which we have found  the
expressions (\ref{4.2}) and (\ref{4.7}).  
Our task here is to make rough estimates of this quantity.

(i)  In II, for hot matter having $T=250 \, {\MeV}$, we took 
$\mu^2/\lambda = 1 \, {\GeV}/{\fm}^2$ 
and, for $L\simeq 10 \, {\fm}$, 
$\tilde v \approx 2.5$ giving 
$$
{\mu^2\tilde v\over \lambda} \approx 0.5 \,  {\GeV}^2/{\fm} \>.
$$  
We get 
$$
p_{\perp W}^2 \simeq 5 \, {\GeV}^2 \, \frac{L}{10 \, {\fm}}
$$ 
from (\ref{3.6}) and, taking $\alpha_s=1/3$, 
$$
- \Delta E \simeq 30 \, {\GeV} \, \left(\frac{L}{10 \, {\fm}}\right)^2
$$ 
for a quark jet.  
These are rather big numbers and one should not take their exact values too seriously. 
However, they do suggest that hot matter may be rather effective in
stimulating radiative energy loss and in broadening the 
$p_\perp$-distribution of high energy partons.
\vskip 10pt

(ii)  For cold matter from (\ref{4.7}), with $C_R=C_F=4/3$, using $\rho = 0.15 \,  
{\fm}^{-3}$ 
one finds 
$$
{\mu^2\tilde v\over \lambda} \approx {1\over 
25}\cdot\alpha_s\cdot[xG(x)]\,{\GeV}^2/{\fm}\>.
$$ 
Taking $\alpha_s=1/2$ and $xG=1$ results in much smaller values
$$
p_{\perp W}^2 \simeq {0.2}\, {\GeV}^2\, \frac{L}{10 \, {\fm}}
$$ 
and
$$
- \Delta E \approx 2\, {\GeV}\, \left(\frac{L}{10 \, {\fm}}\right)^2\>.   
$$
These numbers may be reasonable.  
In particular, 
${dp_{\perp W}^2/ d z} = {\mu^2\tilde v\over \lambda} \approx {1\over 50} \,  
{\GeV}^2/{\fm}$ 
is in agreement with
the $p_\perp$-broadening of the $\mu$-pair spectrum[13] so 
$-{d E/ dz} \approx 0.2\, {\GeV}/{\fm} (L/10\, {\fm})$ may be a sensible,
if small,  
estimate of radiative energy loss for high energy quarks.

\vskip 10pt

(iii) From (\ref{4.2}), with 
$\lambda_{LQS}^2 = 0.05\, {\GeV}^2$, 
one finds 
${\mu^2\tilde v\over \lambda} \approx 0.8\alpha_s(Q^2) \, {\GeV}^2/{\fm}$.  
Taking $\alpha_s(Q^2) = 1/3$ gives
$$
{\mu^2\tilde v\over \lambda} \simeq {0.3}\, {\GeV}^2/{\fm}
$$  
and leads to 
$$
p_{\perp W}^2 \simeq  3\, {\GeV}^2\, \frac{L}{10{\fm}}
$$ 
and 
$$ 
- \Delta E\simeq 15\, {\GeV} \, \left(\frac{L}{10 \, {\fm}}\right)^2\>.
$$  
These numbers seem too large.  
(They would correspond to $x G \approx 20 - 25$ if (\ref{4.7}) is used.) 
The problem here, as alluded to earlier, is that present experiments find a
large $p_\perp$-broadening and energy loss for outgoing partons giving
$\lambda_{LQS}^2\approx 0.05-0.1\, {\GeV}^2$ 
while much smaller numbers are found for the broadening of the $\mu$-pair spectrum.

\vspace{1.5 cm}
\noindent
{\bf\large Acknowledgement}
One of the authors (AM) wishes to thank Dr. Michel Fontannaz and 
the LPTHE-Orsay for hospitality and support during part of the time
that this work was being done.\\
Supported in part by the EEC Programme "Human Capital and Mobility", Network 
"Physics at High Energy Colliders", Contract CHRX-CT93-0357.

\vskip 20pt
\noindent
\appendix \myappendixsection{\strut}
\noindent{\bf Appendix A.\ \ \  Basic equation for jet broadening}    
\noindent

\vskip 6pt
\noindent
Here we give a brief derivation of (\ref{2.7}).  The probability density,
$p_n(U^2,t)$, for the incoming parton to have a transverse momentum squared
$U^2$ at time $t$ after $n$ collisions at times $t_1, t_2, \ldots t_n$ 
and involving fixed transverse momentum transfers $\vec Q_1, \vec Q_2, \ldots  \vec 
Q_n$ 
is \cite{BDPS}
\beql{A.1}
p_n(U^2,t) = f_0\left((\vec U-\sum_{i=1}^n \vec Q_i)^2\right) 
\prod_{\ell=1}^n \left[\,V(Q_\ell^2)
e^{-(t_\ell-t_{\ell-1})}\right]\, e^{-(t-t_n)}\>.
\eeq
Thus,
\beq
\label{A.2}
p_n(U^2,t) = e^{-t} f_0\left((\vec U-\sum_{i=1}^n \vec Q_i)^2\right)\prod_{\ell=1}^n 
V(Q_\ell^2)\>, 
\eeq
where $t_0=0$.

The probability density, $P_n(U^2,t)$, to have momentum $\vec U$ at time  $t$
is obtained by integrating over the $\vec Q_i$ and $t_i$ leading to
\beql{A.3}
P_n(U^2,t) = \int \prod_{\ell=1}^n d^2 \vec Q_\ell \int_0^t d t_n \int_0^{t_n}
dt_{n-1}\cdot \cdot \cdot \int_0^{t_2} d t_1 \, p_n(U^2,t)\>.
\eeq
The momentum distribution $f$  is given by the sum over all
possible number of interactions,
\beql{A.4}
f(U^2,t)=\sum_{n=0}^\infty P_n(U^2,t) = P_0(U^2,t) + \sum_{n=1}^\infty
P_n(U^2,t) \>,
\eeq
where the probability to have no interactions with the medium
between times  0 and  $t$ is
\beql{A.5}
P_0(U^2,t) = f_0(U^2) e^{-t}\>.
\eeq

It is easy to relate $P_n$ to $P_{n-1}$. Inserting (\ref{A.2}) 
into (\ref{A.3}) gives
\beeq\label{A.6}
P_n(U^2,t) &=& \int d^2\vec Q_nV(Q_n^2) \int_0^t d t_n e^{-(t-t_n)}
\prod_{\ell=1}^{n-1} d^2\vec Q_\ell V(\vec Q_\ell^2) 
\int_0^{t_n} d t_{n-1}\cdot \cdot \cdot \int_0^{t_2} dt_1 \nonumber\\
&\cdot& \ e^{-t_n} f_0 \left ( (\vec U-\vec Q_n-\sum_{\ell=1}^{n-1} \vec Q_\ell)^2 
\right ) .
\eeeq
Relabeling $\vec Q_n$ as $\vec Q$ and $t_n$ as $t^\prime$,
\beql{A.7}
P_n(U^2,t) = \int_0^t d t^\prime e^{-(t-t^\prime)} \int d^2\vec Q V(Q^2)
P_{n-1}((\vec U-\vec Q)^2, t^\prime) \>, 
\eeq
which has an obvious probabilistic interpretation.  
Finally, using (A.7) in (A.4) leads to (\ref{2.7}).
\vskip 20pt

\myappendixsection{\strut} 
\noindent{\bf Appendix B. \ \ \ \ Relating elementary scatterings to the gluon 
distribution}

\vskip 10pt

In this appendix we derive (\ref{4.4}).  
Much of our discussion here follows \cite{LQS}.
We  assume a one-gluon exchange
approximation at the outset (see Fig.1), the justification of that assumption
being the  small coupling we find for the coupling of the exchanged gluon to
the incident high energy parton.  
In order to have expressions for the gluon
distribution which conform to conventional notation we suppose that the
incident parton is moving in the negative $z$-direction so that $\vec p=0$ and,
using light-cone variables, $p_- \gg p_+$.  
(In order to fix (approximately) the value of $x$  in (\ref{4.4})
we must  consider the parton propagating through the medium to be virtual
though, in our covariant gauge discussion which follows, we may continue to
assume that the exchanged gluon couples in an eikonal manner.)  
We suppose the nucleon, labeled by $P$, is at rest.  Then  
\beql{B.1}
{d\sigma\over d^2\vec q} \propto g^2\int dq_+ dq_- d^4y (P\vert A_+^a(y)
A_+^a(0)\vert P) \delta((p+q)^2) e^{-iqy} \>,
\eeq
with  $A_\mu^a$ the gluon field with colour index $a$.  
The process is illustrated in
Fig.2 where  0  and   $y$   are the coordinates where the gluon attaches
to the high energy parton in the production amplitude and complex conjugate
amplitude, respectively.

\begin{figure}[h]
\centering
\leavevmode
\epsfxsize 8cm
\epsffile{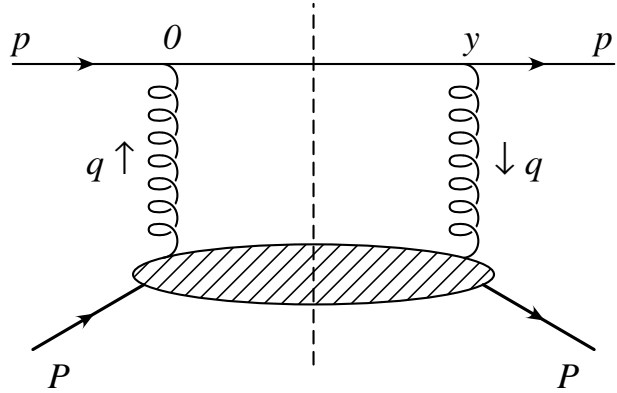}
\caption{\it{Diagrammatic representation of the cross section (B.1).
}}
\label{fig2}
\end{figure}

\noindent With $(q+p)^2 \approx q^2+p^2 + 2 p_-q_+$ the
$q_+-$integration is easily done using the $\delta-$ function in (\ref{B.1}):
\beql{B.2}
{d\sigma\over d^2\vec q} \propto g^2\int dq_-d^4y(P\vert A_+^a(y)
A_+^a(0)\vert P) e^{-iq_+y_--iq_-y_++i\vec q\cdot \vec y}\>,
\eeq
with $q_+ \approx - (p^2+ q^2)/ 2p_- \approx - {p^2/ 2p_-}$ in (\ref{B.2}).  
(We shall see later that $q^2\lsim p^2$.)  
The $q_{-}-$ integration gives a $\delta$-function in $y_+$ 
(because of our large-$p_-$ limit) so we find
\beql{B.3}
{d\sigma\over d^2\vec q} \propto g^2\int d^2\vec y d y_-(P\vert A_+^a(y_+=0,
y_-,\vec y\,) A_+^a(0)\vert P) e^{-iq_+y_-+i\vec q\cdot \vec y}\>.
\eeq
Now multiply both sides of (\ref{B.3}) by $\vec q^{\,2}$.  
The $\vec q^{\,2}$ on the right-hand side of (\ref{B.3}) can be written in terms 
of derivatives of the $A$'s giving
\beql{B.4}
\vec q^{\,2}{d\sigma\over d^2\vec q} \propto g^2 \!\!\int d^2\vec y\,
dy_-(P\vert\,{\partial\over \partial y_i} A_+^a(y_+\!=\!0,y_-,\vec y\,){\partial\over
\partial y_i} A_+^a(0)\,\vert P)\   e^{-iq_+y_-+i\vec q\cdot \vec y} \>.
\eeq
At lowest order in  $g$
\beql{B.5}
F_{i+}^a(y) = {\partial\over \partial y_i} A_+^a(y)-{\partial\over \partial
y_-} A_i^a(y)\>.
\eeq
The second term on the right-hand side of (\ref{B.5}) is small when used
in (\ref{B.4}) because the $y_-$-derivative can be integrated by parts to give a
$q_+$ while the + index on $A_+$ in the first term naturally yields a factor
of $P_+$ when used in (\ref{B.4}).  
(We shall shortly identify $x \equiv q_+/P_+
\approx {-p^2/ 2p_-P_+}\ =\ {-p^2/ s} \ll 1$ with the Bjorken
$x$-variable of the gluon distribution.)  Thus,
\beql{B.6}
\vec q^{\,2} {d\sigma\over d^2\vec q} \propto g^2\int d^2\vec y \,dy_-(P\vert\,
F_{+i}^a(y_+=0,y_-,\vec y\,)F_{+i}^a(0)\,\vert P) e^{-ixP_+y_-+i\vec q\cdot \vec y}.
\eeq
Integrating (\ref{B.6}) over $\vec q^{\,2}$ between 0 and $\mu^2/B^2$ gives
\beql{B.7}
\int_0^{\mu^2/B^2} d^2\!\vec q\, \vec q^{\,2} {d\sigma\over d^2\vec q} \propto
g^2\int_0^\infty\! d\rho J_1(\rho) \int\! dy_-e^{-ixP_+y_-}
(P\vert F_{+i}^a(y_+ =0,y_-,\vec y)
\cdot  F_{+i}^a(0)\vert P)\bigg\vert_{\vec y^2=\rho^2 
\frac{B^2}{\mu^2}}\>, 
\eeq
where we have used $zJ_0(z) = d/dz(zJ_1(z))$.  
Values of $\rho$ on the order of 1 dominate the integral in (\ref{B.7}). 
Therefore in the leading logarithmic approximation we may write
\beq
\label{B.huit}
\int_{0}^{\mu^2/B^2} d{\vec q}^{\ 2} {\vec q}^{\ 2} {d\sigma\over d^2\vec q} \propto 
g^2\! \int dy_-e^{-ixP_+y_-}(P\vert F_{+i}^a(y_+=0,y_-,\vec y\,)F_{+i}^a(0)\vert P)
\bigg\vert_{\vec y^{\,2}=B^2/\mu^2} \>.
\eeq
The integral on the right-hand side of (\ref{B.huit}) is proportional to a
standard expression for the gluon distribution $xG(x, \mu^2/B^2)$ at a momentum
scale $\mu^2/B^2$ \cite{CSS}.  
The $g^2$ in (\ref{B.huit}) (the $\alpha_s$ in (\ref{4.4}))  should also
be evaluated at a scale $\mu^2/B^2$ since this is the limit on the $\vec q^{\,2}$
integration set by the {\em hard}\/ scattering. 
Once we know that the left-hand side
of (\ref{B.huit}) is proportional to   $xG$,  
the constant of proportionality is easily set using (\ref{4.5}) and (\ref{4.6}).

Finally, we come to the value of  $x$   to be used. To determine the approximate
value of  $x$  we must decide what  value of $p^2$ should be used for the
incident parton.  We suppose that the incident parton (jet) is produced in the
nucleus in a hard collision with $\vec p = 0, p_-$ fixed and $p_+$ varying
with the distance from 
  the 
production 
  point 
as $p_+ \simeq 
1/z$.  
(We recall that due to the uncertainty principle one cannot specify $p_+$ 
more accurately in a distance $z$.)  
Then for our elementary scattering
\beql{B.9}
{2p_-\over L} \lsim \vert p^2 \vert \lsim {2p_-\over \lambda_R} \>,
\eeq
where $L$ is the length of the nuclear material and $\lambda_R$ the
mean free path.  
Since
\beql{B.10} 
\vec{q}\,^2 \lsim {\mu^2\over B^2} \sim {L\over 4\lambda_R}\tilde 
v(\lambda_R/L)
\mu^2
\eeq
(see(3.2)),

\beql{B.11}
{\vec q\ ^2\over \vert p^2\vert}\  \lsim\  {\mu^2L^2\tilde v\over
8\lambda_Rp_-},
\eeq

\noindent a quantity which we assume not to be large.   (Note that for $L=L_{cr}$, with
$L_{cr}$ defined in
\citd{BDMPSII}{BDPS}, one has
${\mu^2L^2/ \lambda_Rp_-} \approx 1$.)  Thus, we are justified in dropping $q^2$ 
compared to
$p^2$ in determining $x$ from
$(q + p)^2 = 0$.  

From (\ref{B.9}) we come to the conclusion that $x$ should
be chosen in the range
\beql{B.12}
{1\over M L} \leq\  x\  \leq  {1\over M \lambda_R} \>,
\eeq
with  $M$ the nucleon mass.

\newpage
\def\labelenumi{[\arabic{enumi}]}
\noindent
{\bf\large References}

\begin{enumerate}

\item\label{BDMPSII} R.~Baier, Yu.L.~Dokshitzer, A.H.~Mueller, S.~Peign\'e and 
D.~Schiff,
preprint LPTHE-Orsay 96-34 (May, 1996), hep-ph 9607355.

\item\label{GW} M.~Gyulassy and X.-W.~Wang, \np{420}{583}{94}.

\item\label{Glauber} R.J.~Glauber, in ``Lectures in Theoretical Physics'', 
ed.~W.E.~Brittin and L.G.~Dunham, Interscience, New York, 1959, vol.~1.  

\item\label{LQS} M.~Luo, J.~Qiu and G.~Sterman, \pr{D49}{4493}{94}.

\item\label{Gribov} V.N.~Gribov, \sjnp{9}{369}{69}.

\item\label{BDPS} R. Baier, Yu.L. Dokshitzer, S.~Peign\'e and D.~Schiff,
\pl{345}{277}{95}; \\
R. Baier, Yu.L. Dokshitzer, A.H.~Mueller, S.~Peign\'e and D.~Schiff,
hep-ph 9604327 (February 1996), to  be published in {\it{Nucl.\ Phys.}}\ \underline{B}.

\item\label{Frankfurt} L.L.~Frankfurt, M.I.~Strikman and S.~Liuti,
\prl{65}{1725}{90}.  

\item\label{Eskola} K.J.~Eskola, \np{400}{240}{93}. 

\item\label{Migdal} A.B.~Migdal, \pr{103}{1811}{56}; 
M.L.~Ter-Mikaelian, ``High Energy Elecrtomagnetic Processes in Condensed Media'',
John Wiley and Sons, New York, 1972.

\item\label{BBL} G.T.~Bodwin, S.J.~Brodsky and G.P.~Lepage, 
\prl{47}{1799}{81}. 

\item\label{BH} S.J.~Brodsky and P.~Hoyer, \pl{298}{165}{93}.

\item\label{FC} T.~Fields and M.D.~Corcoran, in the EPS Conference proceedings,
Marseille, France, July 22--28, 1993;  
\prl{70}{143}{93}; \\
R.C.~Moore, et al., \pl{244}{347}{90}; \\
M.D.~Corcoran, et al., \pl{259}{209}{91}.

\item\label{Alde} D.M.~Alde, et al., \prl{66}{2285}{91},
and references therein.

\item\label{LR} E.M.~Levin and M.G.~Ryskin, \sjnp{33}{901}{81}.

\item\label{Levin} E.M.~Levin, hep-ph 9508414 (1995). 

\item\label{CSS} J.C.~Collins, D.E.~Soper and G.~Sterman, 
in ``Perturbative Quantum Chromodynamics'', ed. A.H.~Mueller, 
World Scientific, Singapore, 1990.

\end{enumerate}

\end{document}